\documentclass[12pt]{article}
\usepackage[utf8]{inputenc}
\usepackage[T1]{fontenc}
\usepackage{graphicx}
\usepackage{epstopdf} 
\usepackage{color}
\input{epsf}
\usepackage{lmodern}
\usepackage{amsmath}
\newcommand{\be}{\begin{equation}}
\newcommand{\ee}{\end{equation}}
\newcommand{\bear}{\begin{eqnarray}}
\newcommand{\ear}{\end{eqnarray}}

\begin{document}
\title{On the least uncomfortable journey from $A$ to $B$}
\author{Nivaldo A. Lemos  \\
\small
{\it Instituto de F\'{\i}sica - Universidade Federal Fluminense}\\
\small
{\it Av. Litor\^anea, S/N, Boa Viagem, 
24210-340, Niter\'oi - Rio de Janeiro, Brazil}\\
\small
{\it  nivaldo@if.uff.br}}

\date{\today}

\maketitle

\begin{abstract}

The problem of the ``least uncomfortable journey'' between two locations on a straight line, originally discussed by Anderson {\it et al.} (2016 {\it Am. J. Phys.} {\bf 84} 6905) is revisited. When the  integral of the square of the acceleration is used as a  measure of the discomfort, the problem is shown to be easily solvable by taking the time, instead of the position, as the independent variable. The solution is quite simple and avoids not only complicated differential equations and the computation of cumbersome integrals,  but also the inversion of functions by solving cubic equations. Next, the same problem, but now   with the integral of the square of the jerk as a  measure of the discomfort, is also exactly solved with time as the independent variable and the appropriate boundary conditions, which are derived. It is argued that the boundary conditions imposed on the velocity in Anderson {\it et al.} (2016 {\it Am. J. Phys.} {\bf 84} 6905) are  inappropriate not only because they are not always physically realizable but also  because they do not lead to the minimum discomfort possible.

\end{abstract}

\maketitle

\section{Introduction}

The calculus of variations is one of the most important mathematical tools of theoretical physics. The design of simple yet unorthodox variational problems of physical interest is welcome since they help to acquaint advanced undergraduate or beginning graduate students with several useful techniques of the calculus of variations.  Recently, Anderson, Desaix, and Nyqvist (ADN) \cite{Anderson} discussed the very interesting problem of determining the least uncomfortable way to travel from point $A$ to point $B$ on a straight line, with both the travel time and the distance between the two points fixed.  ADN first minimized the integral of the square of the acceleration, taken as a  measure of the discomfort felt during the journey. Next  they minimized the integral of the square of the jerk as another possible measure  of the discomfort. Less than a year later, a relativistic generalization of the first version  of the problem was published    \cite{Antonelli}.

Frequent acceleration and deceleration  make a trip
uncomfortable, and the discomfort can be reasonably quantified by the integral of the square of the acceleration taken over the duration of the journey. In the case in which  the discomfort is measured by the integral of the square of the acceleration, ADN chose the position $x$ as the independent variable and found the  velocity as a function of the position, $v(x)$, that minimizes the discomfort. 
Their approach is perfectly valid but leads to a rather complicated differential equation, Eq. (9) of ADN, whose solution gives directly $x$ as a function of $v$. The inversion of their Eq. (10) to get $v(x)$ requires the trigonometric solution of a cubic equation. A daunting integral gives $t$ as a function of $x$, whose inversion to yield $x(t)$ is not straightforward either. In Section II we show that the problem is easily solvable by taking the time, instead of the position, as the independent variable. The solution is very easy and avoids not only complicated differential equations and the computation of intimidating integrals,  but also the inversion of functions by solving cubic equations.
  
A high constant acceleration (quite a few $g$-factors) is unpleasant and  even harmful to the human body \cite{Eager}. An abrupt change of a constant acceleration is also a source of distress. Since changes in  acceleration may possibly cause  more discomfort than changes in velocity, it makes sense to define the discomfort in terms of the time rate of change of the acceleration, usually known as jerk \cite{Eager,Schot}. In the case in which the discomfort is measured by the integral of the square of the jerk  one has to cope with a higher-derivative variational problem of a kind  rarely encountered in mechanics. By taking again  the position $x$ as the independent variable, ADN were led to an extremely complicated  higher-derivative functional expressed in terms of $V(x) = v(x)^2$, where $v(x)$ is the velocity as a function of  the position.   Since the ensuing Euler-Lagrange equation for $V(x)$,  their Eq. (19), is too intricate to be exactly solved, ADN resorted to the Rayleigh-Ritz and moment methods in order to find approximate solutions. 

In variational problems, especially those involving higher-order derivatives of the unknown function,  the determination of the appropriate boundary conditions is crucial.  According to 
ADN, the  boundary conditions must be such that ``the initial 
acceleration/deceleration must be zero in order to have finite jerk at the
beginning and end of the journey.'' Consequently, their trial velocity functions were chosen in such a way that the acceleration vanishes at the beginning and at the end of the journey. However, since the journey starts from rest and ends at rest, if the initial acceleration is zero then the initial force is also zero and, in most physical cases,  the velocity will remain equal to zero for all
 time.\footnote{There are exceptions, particularly if the force depends only on time. For instance, if $F(t) = 6t$ the equation of motion ${\ddot x} = F(t)$ with $x(0)=0,  { \dot x}(0) = 0, {\ddot x}(0)= 0$ is solved by $x(t) = t^3$, and it follows that $v(t)= 3t^2$.}  If the force depends on position alone, there can only be a nontrivial motion with both initial velocity and initial acceleration equal to zero if the force is singular, in which case the uniqueness theorem for the solutions of Newton's equations of motion fails \cite{Dhar}. Therefore,  the realization of the boundary conditions  assumed by ADN might require  pathological, unphysical forces. Also, the specification of the initial acceleration is at odds with the basic principles of Newtonian mechanics. These observations motivated us to search for boundary conditions which are more natural to the problem and free from such  physical objections.

In Section III we reexamine the problem of the least uncomfortable journey with the discomfort measured by the integral of the square of the jerk. Time is again taken 
as the independent variable and, by a combination of physical and mathematical arguments,  we derive the physically appropriate boundary conditions. 
Then the exact solution is straightforwardly found and it is shown that it yields for the minimum discomfort a value which is much smaller than the one implied by the ADN boundary conditions. 

Section IV is dedicated to some final remarks.

\section{Discomfort measured by the acceleration}

When  the integral of the square of the acceleration is taken as the  measure of the discomfort, let us show that the problem of the least uncomfortable journey between two locations on a straight line becomes much simpler if one takes the time as the independent variable, instead of the position. Let the coordinate system be so chosen that the departure point $A$ corresponds to  $x= 0$ and the arrival point $B$ corresponds 
to  $x= D$. The vehicle departs from $A$ at $t=0$ and  arrives at $B$
 when $t=T$. The travel time $T$ is  fixed. With the time $t$ as the independent variable, the discomfort functional to be minimized is

\begin{equation}
\label{discomfort-functional}
J[v] = \int_0^T {\dot v}^2 dt 
\end{equation}	
with the boundary conditions
\begin{equation}
\label{boundary-conditions-acceleration}
v(0) = 0, \,\,\,\,\,\, v(T) = 0, 
\end{equation}
and under the isoperimetric constraint
\begin{equation}
\label{isoperimetric-constraint-acceleration}
\int_0^T v dt = D. 
\end{equation}

According to Euler's rule \cite[section 5.3]{Kot}, in order to find the minimizer $v(t)$  one must set up the Euler-Lagrange equation
\begin{equation}
\label{Euler-Lagrange}
\frac{d}{dt}\bigg( \frac{\partial L}{\partial {\dot v}} \bigg) -  \frac{\partial L}{\partial v}= 0 
\end{equation}
with the ``Lagrangian''
 \begin{equation}
\label{Lagrangian-acceleration}
L = {\dot v}^2 + \lambda v, 
\end{equation}
where the Lagrange multiplier $\lambda$ is a constant. If $v$ were the position of a particle, (\ref{Lagrangian-acceleration}) would be  the Lagrangian for a particle with mass
$m=2$ subject to a constant force $\lambda$.

With the ``Lagrangian'' (\ref{Lagrangian-acceleration})
the Euler-Lagrange equation (\ref{Euler-Lagrange}) becomes
\begin{equation}
\label{Euler-Lagrange-explicit}
2{\ddot v} - \lambda = 0 .
\end{equation}
By two successive integrations  the general solution to this equation is  immediately found:
\begin{equation}
\label{Euler-Lagrange-general-solution}
v = A + Bt +  \frac{\lambda}{4} t^2,
\end{equation}
where $A$ and $B$ are constants.
The boundary conditions (\ref{boundary-conditions-acceleration}) yield $A=0$ and $\lambda = -4B/T$. Therefore,
\begin{equation}
\label{Euler-Lagrange-solution-boundary-conditions}
v = BT\bigg(\frac{t}{T} - \frac{t^2}{T^2}\bigg).
\end{equation}
Finally, the isoperimetric constraint (\ref{isoperimetric-constraint-acceleration}) determines the remaining constant  as $B= 6D/T^2$. Thus,  the least uncomfortable journey is achieved by the velocity as the following function of time:
\begin{equation}
\label{v-of-t-least-uncomfortable}
v = \frac{6D}{T}\bigg(\frac{t}{T} - \frac{t^2}{T^2}\bigg).
\end{equation}
The optimal velocity grows steadily from zero when  $t=0$ to its maximum value when $t=T/2$, then  decreases monotonically
  to zero when $t=T$, being a symmetric function about $t=T/2$.
The maximum value of the velocity is
\begin{equation}
\label{maximum-v-acceleration}
v_{max} = \frac{3D}{2T},
\end{equation}	
which is 50\% larger than the average  velocity $v_{ave} = D/T$.

Inasmuch as $x(0)=0$, the position as a function of time is given by
\begin{equation}
\label{x-of-t-least-uncomfortable-implicit}
x(t)   = \int_0^t v(t^{\prime})dt^{\prime} = \frac{6D}{T}\int_0^t \bigg(\frac{t^{\prime}}{T} - \frac{{t^{\prime}}^2}{T^2}\bigg) dt^{\prime},
\end{equation}
whence
\begin{equation}
\label{x-of-t-least-uncomfortable}
x = D \bigg(   3\frac{t^2}{T^2} - 2\frac{t^3}{T^3}\bigg),
\end{equation}
which is equivalent to Eq. (13) of ADN because they picked the coordinate system in such a way that $x(0)=-D$ and $x(T)=D$. 

It is also worth noting that, in the present case, the least uncomfortable journey takes place with constant jerk: $j = {\ddot v} = -12D/T^3$.

The reader should compare the extreme simplicity of our solution with the difficult steps that led to the same result in ADN \cite{Anderson}, whose approach, based on taking $x$ as the independent variable, requires the integration of a complicated differential equation, the trigonometric solution of cubic equations and the computation of nontrivial integrals.

This is a good example of how the choice of independent variable can sometimes make a  variational problem much more tractable.

\section{Discomfort measured by the jerk}

We presently  show that this second version of the  problem can also be explicitly solved in a very simple way once time is adopted as the independent variable and the  physically appropriate boundary conditions are taken into consideration. Once again we choose the coordinate system in such a way that point $A$ corresponds to  $x= 0$ and point $B$ corresponds 
to  $x= D$. 
As before, the vehicle departs  from $A$ at $t=0$ and  arrives at $B$
 when $t=T$. The travel time $T$ is  fixed. With the time $t$ as the independent variable, the discomfort functional to be minimized is
\begin{equation}
\label{discomfort-functional-jerk}
J[v] = \int_0^T {\ddot v}^2 dt 
\end{equation}	
with the boundary conditions
\begin{equation}
\label{boundary-conditions-functional-jerk}
v(0) = 0, \,\,\,\,\,\, v(T) = 0, 
\end{equation}
and under the isoperimetric constraint
\begin{equation}
\label{constraint-functional-jerk}
\int_0^T v dt = D. 
\end{equation}
The process of minimizing the functional (\ref{discomfort-functional-jerk}) under the constraint (\ref{constraint-functional-jerk}) must take into account that,  on physical grounds, no a priori boundary conditions other than (\ref{boundary-conditions-functional-jerk}) should be imposed on $v(t)$. 

In the standard treatment of the problem of minimizing a functional that 
depends on the second derivative of the unknown function, one is naturally 
led to require that both the values of the  unknown function and its derivative be
prescribed at the endpoints \cite[section 4.1]{Kot}. In the present case, however, ${\dot v}(0)$ and 
${\dot v}(T)$  should not be prescribed for the physical reason given in  
the Introduction, namely,  since
the initial and final accelerations are determined by the force acting on the vehicle at
the beginning and at the end of the journey, as a general rule their values cannot be specified  a priori. 
In fact, finding the values of  ${\dot v}(0)$ and 
${\dot v}(T)$ is an integral part of the variational problem. This means that ${\dot v}(0)$ and 
${\dot v}(T)$ must be treated as  arbitrary and, as far as ${\dot v}(t)$ is concerned, we have to deal with a variational 
problem with variable endpoints.
Let us briefly review how to tackle this kind of problem \cite[section 9.1]{Kot}.

\subsection{Variational considerations and boundary conditions}

For a functional of the form
\begin{equation}
\label{higher-derivative-functional-S}
S[v] = \int_0^T L(v, {\dot v}, {\ddot v})dt, 
\end{equation}
the variation 
\begin{equation}
\label{admissible-variations}
v(t) \,\, \longrightarrow \,\, {\bar v}(t) = v(t) + \eta (t) 
\end{equation}
induces the first variation of the functional, defined by
\begin{equation}
\label{variation-functional-higher-derivative}
\delta S = \int_0^T \bigg( \frac{\partial L}{\partial v} \eta + 
\frac{\partial L}{\partial {\dot v}} {\dot \eta} 
+ \frac{\partial L}{\partial {\ddot v}} {\ddot \eta}
\bigg)  dt.
\end{equation}
Two successive integrations by  parts give
\begin{eqnarray}
\label{variation-functional-higher-derivative-integration-parts}
\delta S  & = &\int_0^T \bigg[ \frac{\partial L}{\partial v}  - \frac{d}{dt}\bigg( 
\frac{\partial L}{\partial {\dot v}}\bigg) + \frac{d^2}{dt^2}\bigg( 
\frac{\partial L}{\partial {\ddot v}}\bigg)\bigg] \eta dt \nonumber \\
& & + \bigg[\bigg(\frac{\partial L}{\partial {\dot v}} -  \frac{d}{dt} 
\frac{\partial L}{\partial {\ddot v}}\bigg)\eta \bigg]_0^T   
+ \frac{\partial L}{\partial {\ddot v}} {\dot \eta}\bigg\vert_0^T .
\end{eqnarray}
If the admissible variations are restricted only by the requirement that they satisfy
\begin{equation} 
\label{admissible-variations}
\eta (0) = \eta (T) = 0,
\end{equation}
the variation (\ref{variation-functional-higher-derivative-integration-parts}) reduces to  
\begin{equation}
\label{variation-functional-higher-derivative-integration-parts-reduced}
\delta S = \int_0^T \bigg[ \frac{\partial L}{\partial v}  - \frac{d}{dt}\bigg( 
\frac{\partial L}{\partial {\dot v}}\bigg) + \frac{d^2}{dt^2}\bigg( 
\frac{\partial L}{\partial {\ddot v}}\bigg)\bigg] \eta dt + \frac{\partial L}{\partial {\ddot v}} {\dot \eta}\bigg\vert_0^T .
\end{equation}
The first variation of the functional 
must vanish for {\it all} allowed variations $\eta (t)$, no matter what  the values of ${\dot \eta} (0)$ and ${\dot \eta} (T) $ may be.  Therefore, it must also vanish for those variations 
such that ${\dot \eta} (0) = {\dot \eta} (T) = 0$, for which the boundary term on the right-hand side of (\ref{variation-functional-higher-derivative-integration-parts-reduced}) is zero. In this case, because $\eta$ is otherwise arbitrary,  the fundamental lemma of the calculus of variations \cite[p. 39]{Kot} establishes that  the  condition $\delta S = 0$, which is necessary for a minimum, implies that the coefficient of $\eta$ in the integral on the right-hand side of (\ref{variation-functional-higher-derivative-integration-parts-reduced}) vanishes. Consequently, the minimizer $v(t)$ must satisfy the generalized Euler-Lagrange equation
\begin{equation}
\label{Euler-equation-higher-derivative}
 \frac{\partial L}{\partial v}  - \frac{d}{dt}\bigg( 
\frac{\partial L}{\partial {\dot v}}\bigg) + \frac{d^2}{dt^2}\bigg( 
\frac{\partial L}{\partial {\ddot v}}\bigg) = 0.
\end{equation}
Insertion of this equation into (\ref{variation-functional-higher-derivative-integration-parts-reduced})  further reduces the first variation of the functional $S$  to  
\begin{equation}
\label{variation-functional-higher-derivative-reduced}
\delta S =  \frac{\partial L}{\partial {\ddot v}} {\dot \eta}\bigg\vert_0^T .
\end{equation}
Since the values of ${\dot \eta} (t)$ are arbitrary at $t=0$ and $t=T$, we can choose the variation $\eta (t)$ such that ${\dot \eta} (0) \neq 0, {\dot \eta} (T) = 0$ and vice versa. Therefore, the  condition  $\delta S = 0$ requires that the following boundary conditions be obeyed:
\begin{equation}
\label{variation-functional-higher-derivative-boundary}
\frac{\partial L}{\partial {\ddot v}}\bigg\vert_{t=0} = 0, \,\,\,\,\, \frac{\partial L}{\partial {\ddot v}}\bigg\vert_{t=T} = 0.
\end{equation}
These are {\it natural boundary conditions} \cite[p. 194]{Kot} for the problem with
variable endpoints as regards ${\dot v}(t)$. 

The specification of the appropriate boundary conditions, in addition to the equations of motion, is  an essential part of a physical  problem stated in the form of a variational principle. It is worth noting at this juncture that the
boundary conditions (\ref{variation-functional-higher-derivative-boundary}) are  similar to the ``edge conditions'' that arise in bosonic string theory \cite{Scherk}.

\subsection{The least uncomfortable journey}

Now back to the  problem of the least uncomfortable journey. Since (\ref{boundary-conditions-functional-jerk}) are the only a priori boundary conditions that the velocity is  demanded to obey, the admissible variations are only required to satisfy (\ref{admissible-variations}), and the previous discussion applies. According to Euler's rule \cite[section 5.3]{Kot}, in order to find the minimizer $v(t)$  for the functional (\ref{discomfort-functional-jerk}) under the constraint (\ref{constraint-functional-jerk}), one must set up the generalized Euler-Lagrange equation (\ref{Euler-equation-higher-derivative})
with the ``Lagrangian''
 \begin{equation} 
\label{Lagrangian}
L = {\ddot v}^2 + \lambda v, 
\end{equation}
where the Lagrange multiplier $\lambda$ is a constant. 
With this ``Lagrangian'' 
the generalized Euler-Lagrange equation (\ref{Euler-equation-higher-derivative}) becomes
\begin{equation}
\label{Euler-Lagrange-explicit}
2\frac{d^4 v}{dt^4} + \lambda = 0 .
\end{equation}
From (\ref{Lagrangian}) it follows that the  boundary conditions (\ref{variation-functional-higher-derivative-boundary}) become simply
\begin{equation}
\label{boundary-conditions-jerk}
{\ddot v}(0) =0, \,\,\,\,\, {\ddot v}(T) =0.
\end{equation}

The general solution to  Eq. (\ref{Euler-Lagrange-explicit}) is  immediately written as
\begin{equation}
\label{Euler-Lagrange-general-solution}
v = c_0  + c_1 t + c_2t^2 + c_3t^3 - \frac{\lambda}{48}t^4,
\end{equation}
where $c_0, c_1,  c_2,  c_3$ are constants.
The boundary conditions $v(0)=0$ and ${\ddot v}(0) =0$ yield $c_0=c_2 =0$. The boundary conditions $v(T)=0$ and ${\ddot v}(T) =0$ lead to
\begin{equation}
\label{Euler-Lagrange-general-solution-condition-constants}
c_1 T + c_3T^3 - \frac{\lambda}{48}T^4 =0,\,\,\,\,\, 6c_3T - \frac{\lambda}{4}T^2 =0,
\end{equation}
which are solved by 
\begin{equation}
\label{Euler-Lagrange-general-solution-condition-constants}
\lambda = 24 \frac{c_3}{T}, \,\,\,\,\, c_1 = - \frac{c_3T^2}{2}.
\end{equation}
It follows that
\begin{equation}
\label{Euler-Lagrange-solution-boundary-conditions}
v = c_3T^3\bigg(-\frac{t}{2T} + \frac{t^3}{T^3}  -\frac{t^4}{2T^4}\bigg).
\end{equation}
Finally, the isoperimetric constraint (\ref{constraint-functional-jerk}) determines the remaining constant  as $c_3= - 10D/T^4$. Thus,  the least uncomfortable journey is achieved by the velocity as the following function of time:
\begin{equation}
\label{v-of-t-least-uncomfortable-jerk}
v = \frac{5D}{T}\bigg(\frac{t}{T} - 2\frac{t^3}{T^3} + \frac{t^4}{T^4}\bigg).
\end{equation}
This optimal velocity yields for the discomfort its minimum value:
\begin{equation}
\label{discomfort-functional-jerk-minimum}
J_{min} = \int_0^T {\ddot v}^2 dt = \frac{3600 D^2}{T^6}\int_0^T \bigg(\frac{t}{T} - \frac{t^2}{T^2}\bigg)^2 dt = 120 \frac{D^2}{T^5}.
\end{equation}	

With the change of variable $\tau = t -T/2$,   the right-hand side of (\ref{v-of-t-least-uncomfortable-jerk}) does not contain odd powers of $\tau$. Therefore, the  velocity that minimizes the discomfort is  a symmetric function about $t=T/2$ that grows steadily from zero at $t=0$ to a maximum at $t=T/2$, then  decreases monotonically
 from the maximum at $t=T/2$ to zero at $t=T$.
The maximum value of the velocity is
\begin{equation}
\label{maximum-v-jerk}
v_{max} = \frac{25D}{16T},
\end{equation}	
which is about 56\% larger than the average  velocity $v_{ave} = D/T$.	This means that, when the discomfort is measured by the acceleration, the least uncomfortable journey is slightly smoother than when the discomfort is measured by the jerk, as illustrated in Fig. 1.
\begin{figure}[h!] 
	\centering
	\includegraphics[width=0.45\linewidth]{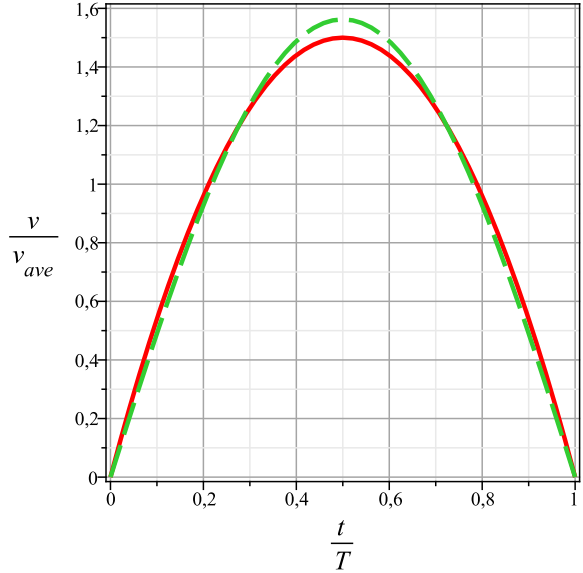}
	\caption[Circle]{The solid  line represents the velocity given by Eq. (\ref{v-of-t-least-uncomfortable}), whereas the dashed line  stands for the velocity given by Eq. (\ref{v-of-t-least-uncomfortable-jerk}), with $v_{ave}=D/T$.}
\label{comparison-graph}
\end{figure}

By the way, it should be noted that, by approximate methods based on velocity trial functions enforcing physically doubtful boundary conditions, Anderson, Desaix, and Nyqvist  found \cite{Anderson2}
\begin{equation}
\label{maximum-v-jerk-ADN}
v_{max}^{\mbox{\tiny ADN}} = 1.84 \frac{D}{T},
\end{equation}	
which is more than  $17\%$ above the exact value (\ref{maximum-v-jerk}). The discrepancy in the predicted minimum value for the discomfort is much more pronounced,  as we proceed to show. Starting from (\ref{Euler-Lagrange-general-solution}), the implementation of the isoperimetric constraint (\ref{constraint-functional-jerk}) and  of the ADN boundary conditions, namely $v(0)=v(T)=0$ as well as ${\dot v}(0) = {\dot v}(T) =0$, leads to the supposedly optimal velocity
\begin{equation}
\label{v-of-t-least-uncomfortable-jerk-ADN}
v^{\mbox{\tiny ADN}} = \frac{30D}{T}\bigg(\frac{t^2}{T^2} - 2\frac{t^3}{T^3} + \frac{t^4}{T^4}\bigg).
\end{equation}
The discomfort associated with this velocity is  
\begin{equation}
\label{discomfort-functional-jerk-minimum-ADN}
J_{min}^{\mbox{\tiny ADN}} = \int_0^T ({\ddot v}^{\mbox{\tiny ADN}})^2 dt = \frac{3600 D^2}{T^6}\int_0^T \bigg(1- 6\frac{t}{T} + 6 \frac{t^2}{T^2}\bigg)^2 dt = 720 \frac{D^2}{T^5}.
\end{equation}
This is six times as large as the true minimum  (\ref{discomfort-functional-jerk-minimum}).


In Fig. 2 the optimal velocity given by Eq. (\ref{v-of-t-least-uncomfortable-jerk}) is compared with the  velocity (\ref{v-of-t-least-uncomfortable-jerk-ADN}) derived from the inappropriate ADN boundary conditions. The ADN boundary conditions lead  not only to a much larger discomfort but also to a maximum velocity which is 20\% above the exact value.
\begin{figure}[h!] 
	\centering
	\includegraphics[width=0.45\linewidth]{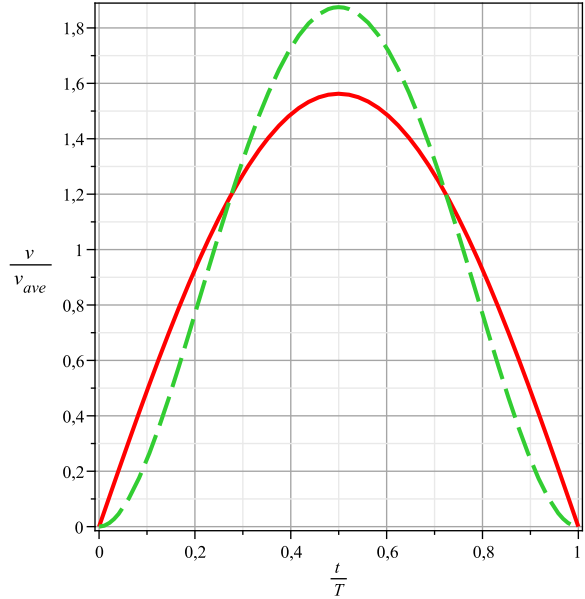}
	\caption[Circle]{The solid  line represents the true optimal velocity given by Eq. (\ref{v-of-t-least-uncomfortable-jerk}), whereas the dashed line stands for the  velocity (\ref{v-of-t-least-uncomfortable-jerk-ADN}) derived from the inappropriate ADN boundary conditions, with $v_{ave}=D/T$.}
\label{comparison-graph}
\end{figure}

 Returning to the exact solution (\ref{v-of-t-least-uncomfortable-jerk}), the acceleration is given by
\begin{equation}
\label{a-of-t-least-uncomfortable}
a = {\dot v} = \frac{5D}{T^2}\bigg(1 -6\frac{t^2}{T^2} + 4\frac{t^3}{T^3}\bigg).
\end{equation}
As physically expected, the acceleration takes a positive value ($a_0=5D/T^2$) at the beginning of the journey ($t=0$) and decreases monotonically to a negative value ($a_T=-5D/T^2$) at the end of the journey ($t=T$); it goes  through zero and changes sign  at $t=T/2$, being  an odd function with respect to $t=T/2$. Finally, the jerk is given by  
\begin{equation}
\label{jerk-of-t-least-uncomfortable}
j = {\ddot v} = -\frac{60D}{T^3}\bigg(\frac{t}{T}  - \frac{t^2}{T^2}\bigg).
\end{equation}
The jerk is a negative  function, symmetric about $t=T/2$, that  vanishes at the beginning and  the end of the journey;  its minimum value $j_{min}= -15D/T^3$ is reached at $t=T/2$.


The position is given by 
\begin{equation}
\label{x-of-t-least-uncomfortable-implicit-jerk}
x(t)   = \int_0^t v(t^{\prime})dt^{\prime} = \frac{5D}{T}\int_0^t \bigg(\frac{t^{\prime}}{T} - 2\frac{{t^{\prime}}^3}{T^3} + \frac{{t^{\prime}}^4}{T^4}\bigg) dt^{\prime},
\end{equation}
whence
\begin{equation}
\label{x-of-t-least-uncomfortable-jerk}
x = \frac{D}{2} \bigg( 5\frac{t^2}{T^2} - 5\frac{t^4}{T^4} + 2\frac{t^5}{T^5}\bigg).
\end{equation}
It is now clear why the variational problem for $v(x)$ is so difficult. Finding
$v(x)$ requires the solution
of the fifth-degree algebraic equation (\ref{x-of-t-least-uncomfortable-jerk}) for $t$ in order to determine $t(x)$ for its substitution into Eq. (\ref{v-of-t-least-uncomfortable-jerk}), which is very hard to do. Expressing the solution for $t(x)$ would require the use of the not-so-well-known Jacobi theta functions \cite{quintic}.

\section{Final Remarks}

The problem of the least uncomfortable journey on a straight line makes it clear that  the choice of independent variable may be capable of making a  variational problem much more easily solvable. It   also provides a nice example of a physical variational problem involving a  functional that depends on derivatives of  order higher than the first. As we have seen, in such cases  the identification of the appropriate boundary conditions calls for a careful combination of physical and mathematical arguments. 

The relativistic generalization of the least uncomfortable journey with the discomfort measured by the acceleration cannot be solved in terms of elementary functions \cite{Antonelli}. Although it might be of interest to formulate a relativistic generalization of the same  problem with the discomfort measured by the jerk, there is not much hope that an  exact solution in terms of elementary functions can be found.

\subsection*{Acknowledgment}
The author is thankful to the anonymous reviewers whose criticisms, corrections and suggestions have contributed to a significant improvement of the original manuscript. 



 
 \bibliographystyle{unsrt}

\begin{thebibliography}{99}
\bibitem{Anderson2} See the line right below Eq. (36) in Ref. 1, where  $2D$ is taken for the distance between points $A$ and $B$.
\end{thebibliography}

\end{document}